\documentclass[10pt, conference]{IEEEtran}

\usepackage[english]{babel}
\usepackage[pdftex]{graphicx}

\usepackage{hyperref}
\usepackage[cmex10]{amsmath}
\usepackage{amsmath,amssymb,amsfonts}
\usepackage{booktabs}
\usepackage{xcolor}
\usepackage{multirow}

\font\myfont=cmr10 at 14pt
\title{{\myfont Malware Detection at the Edge with Lightweight LLMs: A Performance Evaluation }}

\newcommand{\linebreakand}{%
  \end{@IEEEauthorhalign}
  \hfill\mbox{}\par
  \mbox{}\hfill\begin{@IEEEauthorhalign}
}

\begin{document}

\author{
\IEEEauthorblockN{Christian Rondanini\IEEEauthorrefmark{1},
Barbara Carminati\IEEEauthorrefmark{1},
Elena Ferrari\IEEEauthorrefmark{1},
Antonio Gaudiano\IEEEauthorrefmark{1}
},
\IEEEauthorblockA{\IEEEauthorrefmark{1}DiSTA, University of Insubria, Italy\\
\{christian.rondanini, barbara.carminati, elena.ferrari, agaudiano\}@uninsubria.it}
\and
% \IEEEauthorblockN{Ashish Kundu\IEEEauthorrefmark{2},
% Akshay Jajoo\IEEEauthorrefmark{2}}
% \IEEEauthorblockA{\IEEEauthorrefmark{2}Cisco Research, USA\\
% \{ashkundu, ajajoo\}@cisco.com}
\IEEEauthorblockN{Ashish Kundu\IEEEauthorrefmark{2}}
\IEEEauthorblockA{\IEEEauthorrefmark{2}Cisco Research, USA\\
ashkundu@cisco.com}
}

\maketitle

\begin{abstract}
The rapid evolution of malware attacks calls for the development of innovative detection methods, especially in resource-constrained edge computing. Traditional detection techniques struggle to keep up with modern malware's sophistication and adaptability, prompting a shift towards advanced methodologies like those leveraging Large Language Models (LLMs) for enhanced malware detection. However, deploying LLMs for malware detection directly at edge devices raises several challenges, including ensuring accuracy in constrained environments and addressing edge devices' energy and computational limits.
To tackle these challenges, this paper proposes an architecture leveraging lightweight LLMs' strengths while addressing limitations like reduced accuracy and insufficient computational power. 
To evaluate the effectiveness of the proposed lightweight LLM-based approach for edge computing, we perform an extensive experimental evaluation using several state-of-the-art lightweight LLMs.
%(such as distilBERT \cite{sanh2019distilbert} and TinyBERT \cite{jiao2019tinybert}), 
 We test them with several publicly available datasets specifically designed for edge and IoT scenarios and different edge nodes with varying computational power and characteristics.
\end{abstract}

\begin{IEEEkeywords}
Malware;  Edge Computing;  Large Language Model; Security
\end{IEEEkeywords}

\section{Introduction}\label{sec:intro}

In recent years, the rapid evolution of malware attacks has necessitated the development of innovative approaches for their detection, particularly in the resource-constrained edge computing domain. While foundational, traditional detection techniques have struggled to keep pace with modern malware's increasing sophistication and adaptability. This has prompted a shift towards exploring advanced methodologies, including using lightweight Large Language Models (LLMs), to enhance malware detection capabilities in edge environments.

Among the various lightweight LLMs, models such as DistilBERT, TinyBERT, Llama-3.2-1B, DistilGPT-2, and TinyT5 have emerged as promising solutions. These models leverage techniques such as distillation and pruning to significantly reduce their size and computational requirements, making them more suitable for edge-devices deployment. Despite their smaller footprint, these models retain much of their larger counterparts' pattern recognition and contextual understanding capabilities, allowing them to process and analyze complex, unstructured data streams effectively. In the context of malware detection, they offer the potential for improved accuracy, real-time adaptability, and continuous learning while addressing the strict energy, storage, and computational constraints of edge computing environments.

However, deploying LLMs for malware detection in edge computing is not without challenges. Key issues include:
\begin{itemize}
 \item {\bf Model performance}: maintaining high accuracy under the constraints of edge environments remains a significant hurdle. While LLMs excel in natural language understanding and pattern recognition, their generalizability across diverse edge scenarios is often limited, particularly when faced with malware’s dynamic and adaptive nature \cite{friha2024llm}.

  \item {\bf Susceptibility to bias}: overfitting during fine-tuning can make LLMs prone to biases, potentially limiting their ability to detect novel malware threats. To mitigate these risks, robust generalization requires diverse training datasets and advanced regularization techniques.

\item {\bf Feasibility of deployment}: LLMs' computational and energy demands challenge their practicality on resource-constrained edge devices. Addressing this requires further advancements in lightweight model design through techniques such as quantization, knowledge distillation, and efficient architecture optimization.

\end{itemize}
To tackle these challenges, this paper proposes an architecture that leverages the strengths of lightweight LLMs while addressing their limitations in edge computing scenarios, such as reduced accuracy and insufficient computational power. The proposed architecture enhances malware detection capabilities, improves adaptability, and supports continuous learning in resource-constrained.

To evaluate the effectiveness of the proposed lightweight LLM-based approach for edge computing, we perform an extensive experimental evaluation using several state-of-the-art lightweight LLMs (such as distilBERT \cite{sanh2019distilbert} and TinyBERT \cite{jiao2019tinybert}).

We run our experiments on different publicly available datasets, specifically designed for edge and IoT scenarios, including Edge-IIoTset \cite{Ferrag_EdgeIOT}, X-IIoTID \cite{AlHawawreh2022}, TON-IoT \cite{moustafa2021new}, and the CIC IoT Dataset 2023 \cite{Neto2023CICIoT2023}. These datasets provide a rich variety of real-world network traffic and malware samples, enabling a comprehensive assessment of the models’ performance in detecting sophisticated threats.
Finally, we test the feasibility of the proposed approach on edge nodes with different computational power and characteristics.
To the best of our knowledge, we are not aware of other proposals performing such an extensive performance evaluation on the deployment of lightweight LLMs in the edge scenario.

The remainder of this paper is organized as follows. Section \ref{sec:background} provides background information on LLM and relevant datasets. Section \ref{sec:applications} surveys related work. Section \ref{sec:challenges+architecture} introduces our proposed architecture.  The results of our extensive performance evaluation are presented in Section \ref{sec:expresults}, whereas Section \ref{sec:conclusions} concludes the paper and outlines future research directions.

\section{Background}\label{sec:background}

In this section, we first present the key elements of LLMs (Section \ref{sub:models}).  Then, we illustrate the public datasets we used in our experimental evaluation (Section  \ref{sub:dataset}).

\subsection{Large language models}\label{sub:models}

The earliest language models were developed in the early 1990s \cite{zhao2023surveyLLM}, using basic machine learning approaches and simpler probabilistic models, such as n-grams \cite{brown1992class} that estimated word probabilities based on previous context.
A significant advancement in the field was the introduction of neural networks, such as recurrent neural networks (RNNs) (e.g., the RNNLM \cite{mikolov2011strategies}) and their variations, such as long short-term memory (LSTM) \cite{sutskever2014sequence}. 
These models became popular for a variety of natural language applications, including text creation, machine translation, and text categorization, since they provide improved text sequence handling capabilities. 
Nevertheless, they still struggled to effectively convey long-term interdependence in text.

% %3Pre-trained language models (PLM)
However, it was 2017 that marked a significant breakthrough in natural language processing (NLP), with the introduction of the transformer architecture by Vaswani et al.  \cite{vaswani2017attention}. 
This innovation enabled parallel processing while effectively capturing long-range dependencies in data, revolutionizing the field.
Its self-attention mechanisms and feed-forward neural networks allow models to assign varying weights to words in an input sequence based on their contextual relevance. 
Additionally, multi-head attention extends self-attention by enabling the model to simultaneously analyze multiple aspects of the data. 
This capability allows models to dynamically adjust the importance of different words within a context while simultaneously processing input sequences in parallel, leading to significant improvements in both efficiency and performance. 
The transformer architecture includes encoder and decoder modules enabled by self-attention mechanisms.
Thanks to self-attention and feedfoward, the encoder processes the input sentences and produces a representation of it. In contrast, the decoder generates a sequence with respect to the encoder representation (if any) and produces a prediction of the next token, again thanks to self-attention and cross-attention.

% % 3-4  Large language models (LLM)
The advent of the transformer architecture and the availability of powerful yet affordable hardware led to the era of Large Language Models (LLMs).
Most LLMs today are based on the transformer architecture, and they mainly differ by the number of parameters used, with those exceeding one billion parameters categorized as large pre-trained language models (PLMs).
However, at the time of writing, there is no formal agreement within the academic community on the minimum parameter threshold for LLMs, as model capacity is closely tied to the size of the training dataset and the computational resources utilized.

To categorize existing LLMs, in what follows, we adopt the framework proposed by Pan et al. \cite{Unifying}, which organizes them into three main categories: encoder-only, encoder-decoder, and decoder-only models.

The encoder-only models use only an encoder to encode sentences and understand the relationships between words.
Encoder-only models are mainly designed for tasks that involve understanding, analyzing, or classifying text. A notable example in this category is BERT (Bidirectional Encoder Representations from Transformers) \cite{devlin2018bert}, introduced by Google in 2018.
In contrast, decoder-only models only adopt the decoder module to generate target output text. 
These models are mainly designed for generative tasks, such as the generation of code or text. A notable example in this category is the Generative Pre-trained Transformer (GPT)  model \cite{gptcore}. 
Lastly, encoder-decoder models combine the capabilities of previously described models to handle complex tasks, such as machine translation and text-to-text transformation. 
A notable example in the encoder-decoder category is T5 (Text-To-Text Transfer Transformer)\cite{raffel2020exploring} by Google.

% 5 model details 
%In this paper, we select the most representative models from each category for evaluation.  
%Additionally, given edge devices' computational and memory limitations, deploying full-scale LLMs on edge devices poses significant challenges.
%Thus, we focus on lightweight variants of the selected LLM models to achieve high malware detection accuracy while maintaining efficiency and responsiveness.

A further step in LLM evolution is given by the growing emphasis on developing light LLM models, also called small language models (SLMs) \cite{xu2024survey}. These models are designed to require significantly reduced computational resources while still delivering high accuracy and performance levels. Techniques such as model pruning, quantization, knowledge distillation, as well as efficient architecture design are commonly employed to achieve a balance between computational requirements and achieved accuracy.

%DistillBERT & Tinybert
Examples of SLMs are DistilBERT \cite{sanh2019distilbert} and TinyBERT \cite{jiao2019tinybert}, which are respectively a distilled and tiny version of the BERT model.
DistilBERT retains 97\% of BERT’s original performance while being 40\% smaller and 60\% faster. 
This substantial reduction in size (i.e., a lesser number of parameters) and inference time is achieved through a process called knowledge distillation, where a smaller student model learns to replicate the outputs of a larger teacher model. 
DistilBERT’s encoder-only architecture, focused on bidirectional contextual understanding, makes it especially effective for classification tasks, such as distinguishing between malicious and benign activities. 

Unlike distilled models, tiny models are not derived from larger ones but are explicitly built to be lightweight and efficient.
TinyBERT builds on the BERT architecture with a more aggressive optimization strategy, employing a two-stage distillation process to maximize both speed and memory efficiency. 
This results in a model significantly smaller and faster than its predecessor, yet capable of preserving critical linguistic features. 
TinyBERT is particularly advantageous for edge-based malware detection in scenarios where rapid, real-time analysis of contextual data is required, such as identifying anomalous patterns in network traffic or system logs.

%GPT2 & TinyLLAMA
For decoder-only models, distilGPT2 \cite{sanh2019distilbert} and Llama 3.2 1B\footnote{https://huggingface.co/meta-llama/Llama-3.2-1B-Instruct} are popular distilled versions
of the GPT  model \cite{gptcore}, and   LLama (Large Language Model Meta AI) \cite{touvron2023llama}, respectively.
DistilGPT2 is a streamlined version of the GPT-2 architecture, designed primarily for generative tasks but adaptable for classification when appropriately fine-tuned. Unlike encoder-only models, DistilGPT2  decoder-only structure is suitable for tasks requiring a focus on sequential dependencies, such as the detection of obfuscated or polymorphic malware encoded in text or script files. 
By reducing the computational overhead of the original GPT-2 model, DistilGPT2 can be an efficient solution for edge devices that require nuanced and context-sensitive anomaly detection. 

%https://ai.meta.com/blog/llama-3-2-connect-2024-vision-edge-mobile-devices/
Llama 3.2 1B \cite{dubey2024llama} represents a distilled version of the LLaMA architecture optimized for inference on edge devices.
Through techniques such as pruning, quantization, and architectural simplifications, Llama 3.2 1B achieves significant efficiency gains while maintaining much of the performance and versatility of its larger counterpart. 
Its design focuses on delivering high throughput and low latency, making it well-suited for resource-constrained environments that demand rapid and accurate threat analysis.
Llama 3.2 1B excels in processing complex input patterns and can be particularly effective for identifying advanced persistent threats (APTs) or zero-day exploits.

Lastly,  for the encoder-decoder category,  TinyT5 \cite{tay2021scale} is a distilled version of  Google  T5 \cite{raffel2020exploring}.
By leveraging knowledge distillation, TinyT5 achieves a smaller model size and faster inference times, making it highly suitable for deployment on edge devices with limited computational and memory resources.
This lightweight model is optimized to handle a wide range of natural language processing tasks in a text-to-text format, maintaining strong performance across benchmarks.

Table \ref{tab:LLMsmodels} summarizes the main characteristics of the above discussed models. 
In the table, Vocabulary size is the number of unique tokens (words or chars) in the model's vocabulary, which determines the size of the input/output layers.
Hidden layers express the number of intermediate layers between the input and output layers.
Hidden size denotes the number of neurons in each hidden layer.
The number of heads indicates the number of attention heads from the transformer architecture.

% \begin{table}[!htbp]
% \resizebox{\textwidth}{!}{%
% \begin{tabular}{@{}|l|c|l|l|l|l|c|c|@{}}
% \toprule
% \multicolumn{1}{|c|}{\multirow{2}{*}{\textbf{Model}}} & \multirow{2}{*}{\textbf{Distilled from}} & \multicolumn{1}{c|}{\multirow{2}{*}{\textbf{\begin{tabular}[c]{@{}c@{}}Hidden\\ size\end{tabular}}}} & \multicolumn{1}{c|}{\multirow{2}{*}{\textbf{\# Layers}}} & \multicolumn{1}{c|}{\multirow{2}{*}{\textbf{\begin{tabular}[c]{@{}c@{}}Vocab\\ size\end{tabular}}}} & \multicolumn{1}{c|}{\multirow{2}{*}{\textbf{\# Parameters}}} & \multirow{2}{*}{\textbf{Speedup}} & \multirow{2}{*}{\textbf{Use case}} \\
% \multicolumn{1}{|c|}{} &  & \multicolumn{1}{c|}{} & \multicolumn{1}{c|}{} & \multicolumn{1}{c|}{} & \multicolumn{1}{c|}{} &  &  \\ \midrule
% distillGPT2 & GPT-2 & 768 & 6 & 50257 & 81,912,576 & 2x faster & Text generation \\ \midrule
% distillBERT & BERT-base & 768 & 6 & 30522 & 66,362,880 & 60\% faster & NLP (classification, QA) \\ \midrule
% tinyBERT & BERT-base & 312 & 4 & 30522 & 14,350,248 & 80\% faster & NLP (lightweight BERT) \\ \midrule
% Llama-3.2-1B & Llama 3 & 2048 & 16 & 128256 & 1,235,814,400 & N/A & General-purpose LLM \\ \midrule
% tinyT5 & T5-base & 256 & 4 & 32128 & 15,570,688 & 2x faster & Text-to-text (summarization, translation) \\ \bottomrule
% \end{tabular}%
% }

\begin{table*}[!htbp]
\resizebox{\textwidth}{!}{%
\begin{tabular}{@{}|l|c|c|c|c|l|l|c|c|@{}}
\toprule
\multicolumn{1}{|c|}{\multirow{2}{*}{\textbf{Model}}} & \multirow{2}{*}{\textbf{Distilled from}} & \multirow{2}{*}{\textbf{\begin{tabular}[c]{@{}c@{}}Hidden\\ size\end{tabular}}} & \multicolumn{1}{l|}{\multirow{2}{*}{\textbf{\# Heads}}} & \multirow{2}{*}{\textbf{\# Layers}} & \multicolumn{1}{c|}{\multirow{2}{*}{\textbf{\begin{tabular}[c]{@{}c@{}}Vocab\\ size\end{tabular}}}} & \multicolumn{1}{c|}{\multirow{2}{*}{\textbf{\# Parameters}}} & \multirow{2}{*}{\textbf{Speedup}} & \multirow{2}{*}{\textbf{Use case}} \\
\multicolumn{1}{|c|}{} &  &  & \multicolumn{1}{l|}{} &  & \multicolumn{1}{c|}{} & \multicolumn{1}{c|}{} &  &  \\ \midrule
distilGPT2 & GPT-2 & 768 & 12 & 6 & 50257 & 81,912,576 & 2x faster & Text generation \\ \midrule
distilBERT & BERT-base & 768 & 12 & 6 & 30522 & 66,362,880 & 60\% faster & NLP (classification, QA) \\ \midrule
TinyBERT & BERT-base & 312 & 12 & 4 & 30522 & 14,350,248 & 80\% faster & NLP (lightweight BERT) \\ \midrule
Llama-3.2-1B & Llama 3 & 2048 & 32 & 16 & 128256 & 1,235,814,400 & N/A & General-purpose LLM \\ \midrule
TinyT5 & T5-base & 256 & 4 & 4 & 32128 & 15,570,688 & 2x faster & \begin{tabular}[c]{@{}c@{}}Text-to-text (summarization,\\  translation)\end{tabular} \\ \bottomrule
\end{tabular}%
}

   \vspace{+0.1em}
   \caption{SLMs characteristics and use cases. 
 }\label{tab:LLMsmodels}
\end{table*}

\subsection{Datasets} \label{sub:dataset}

The selection of datasets is a key factor in training and evaluating LLMs for malware detection in edge environments.
In \cite{rondanini2024large}, we surveyed a comprehensive list of public datasets that can be exploited for testing LLM-based malware detection techniques in the edge computing scenario.
From these datasets, we have selected those that focus on network traffic and are, to the best of our knowledge, the most appropriate for our testing activities, as they offer a good size, age and feature distribution combination.
A detailed overview of the considered datasets is provided in the following, whereas  Table \ref{tab:Dataset} summarizes their main characteristics.

\begin{itemize}

    \item \textbf{X-IIoTID.}
    This dataset was created at the University of New South Wales (UNSW), Canberra\cite{AlHawawreh2022}. It
    constitutes a connectivity- and device-agnostic benchmark for the evaluation and training of machine learning (ML) and deep learning (DL) intrusion detection systems (IDS) within the Internet of Things (IoT) and Industrial IoT (IIoT) environments. 
    The dataset was created in a testbed mimicking the Industrial Internet Reference Architecture (IIRA). 
    This architecture integrates various industrial and IoT devices, communication protocols, cloud services, and attack simulation tools.
    The dataset includes 820,834 records, comprising 421,417 normal and 399,417 malicious instances, with 65 features extracted from network traffic, device resources, and logs associated with devices and alerts. 
    The captured traffic employs a variety of protocols, such as Modbus, MQTT, TCP, CoAP, and SMTP, which reflect the heterogeneous characteristics of IIoT ecosystems.
    Malicious records were systematically generated using three established frameworks: the Cyber Kill Chain (CKC), the Malware Lifecycle Chain (MALC), and the MITRE ATT\&CK framework.
    The simulated attack processes were carefully designed to span multiple stages of a typical cyberattack lifecycle, facilitating a comprehensive assessment of IDS effectiveness. 
    These stages encompass reconnaissance, weaponization, exploitation, lateral movement, command and control (C2), data exfiltration, tampering, cryptographic ransomware deployment, and ransomware-induced denial of service (Ransom DoS).

    \item \textbf{Edge-IIoTset}.
    The Edge-IIoTset \cite{Ferrag_EdgeIOT} dataset, with a size of 1.2 GB, represents a well-calibrated balance between data richness and computational manageability.
    Tailored specifically for edge computing, it captures diverse IIoT traffic patterns and an extensive range of malicious activities, offering a broader threat landscape compared to smaller datasets, such as X-IIoTID \cite{AlHawawreh2022}.
    Edge-IIoTset includes detailed metadata, facilitating a deeper understanding and detection of evolving malware tactics.
    Its medium-scale size supports advanced feature engineering while remaining suitable for lightweight models, ensuring compatibility with resource-constrained edge devices.
    The dataset comprises both benign and malicious traffic samples, encompassing a variety of attack types such as Distributed Denial-of-Service (DDoS), botnets, and ransomware. It also contains labelled data with network flow characteristics, system logs, and behavioural indicators, enabling comprehensive analysis of IIoT security threats.

    \item \textbf{TON-IoT.}
    The TON-IoT dataset, released in 2021 by Moustafa et al. \cite{moustafa2021new}, stands as one of the most comprehensive datasets for IoT-based threat detection, offering a detailed representation of network environments through diverse data sources and modalities.
    Designed within a realistic network setting, it captures an extensive range of attack scenarios, including DoS, DDoS, Man-in-the-Middle, ransomware, and password cracking.
    Data collection was carried out across multiple streams, encompassing network traffic, operating system logs, and telemetry data from IoT sensors, providing a holistic foundation for malware detection.
    The dataset comprises 223 million records with 44 features, supporting a wide variety of analytical approaches.
    Its inclusion of multimodal data enables cross-layer analysis and the integration of multiple data streams, which are essential for developing advanced detection models tailored to complex and multi-vector attacks. 
    The size and versatility of TON-IoT make it particularly valuable for training models designed for edge computing deployments, addressing the challenges of real-world IoT environments while enhancing the detection of sophisticated cyber threats. 

    \item \textbf{CIC IoT Dataset 2023.} 
    The CIC IoT dataset \cite{Neto2023CICIoT2023} is the largest discussed in this paper and spans 12.8 GB.
    It provides a detailed representation of IoT network traffic and malicious behaviours, including botnet activities, ransomware, and stealthy fileless malware. 
    Collected from 105 IoT devices, the dataset encompasses approximately 548 GB of raw traffic data, capturing 33 distinct attack types, such as DDoS, DoS, Reconnaissance, Web-based attacks, Brute Force, Spoofing, and Mirai-based threats.
    Each of its 46.69 million records includes 47 features, such as flow duration, protocol type, packet size (minimum, average, and maximum), and packet capture timestamps.
    The dataset's comprehensive coverage and detailed labelling make it ideal for training high-capacity models.
    The inclusion of diverse attack scenarios and extensive features ensures that models trained on it can generalize effectively to a wide range of IoT security contexts.
    However, the dataset's large size poses challenges for direct edge deployment, making it more suitable for pre-training models in centralized environments.
    These models can then be distilled or fine-tuned for deployment on edge devices, enabling effective detection of complex and multi-vector attacks while addressing the constraints of resource-limited IoT environments.  

\end{itemize}

\begin{table*}[!htbp]
\resizebox{\textwidth}{!}{%
\begin{tabular}{@{}|l|l|c|c|c|c|l|l|l|@{}}
\toprule
\multicolumn{1}{|c|}{\textbf{Name}} & \multicolumn{1}{c|}{\textbf{Description}} & \textbf{Size (GB)} & \textbf{Records} & \textbf{Attack types} & \textbf{\# of Features} & \multicolumn{1}{c|}{\textbf{Source}} & \multicolumn{1}{c|}{\textbf{Year}} & \multicolumn{1}{c|}{\textbf{Format}} \\ \midrule
X-IIoTID & \begin{tabular}[c]{@{}l@{}}It contains attackers behaviors \\ on Industrial IoT devices\end{tabular} & 0.38 & 0.82M & IG, IT, DS & 65 & Real devices & 2022 & CSV \\ \midrule
EdgeIIoTset & Collection of IoT / IIoT device data & 1.2 & 72M & IG, DS & 46 & Simulated & 2022 & PCAP, CSV \\ \midrule
TON-IoT & \begin{tabular}[c]{@{}l@{}}Collection of network traffic data \\ for IoT environment\end{tabular} & 3.68 & 223M & IT, DS & 44 & \begin{tabular}[c]{@{}l@{}}Real-world \\ and simulated\end{tabular} & 2021 & \begin{tabular}[c]{@{}l@{}}CSV, PCAP,\\ BLG, Logs\end{tabular} \\ \midrule
CIC IoT 2023 & A real-time dataset of large-scale attacks in IoT & 12.8 & 46.69M & IG, IT, DS & 47 & \begin{tabular}[c]{@{}l@{}}Real-world devices, \\ Simulation\end{tabular} & \multicolumn{1}{c|}{2023} & CSV, PCAP \\ \bottomrule
\end{tabular}%
}

\vspace{+0.1em}
\caption{Public datasets suitable for edge computing malware detection. Information Gathering (IG), Information Theft (IT), Denial of Service (DS), %Normal (N), Attack Type (AT), Benign (B), Malicious (M),
Million Records(M)}

\label{tab:Dataset}
\end{table*}

\section{Related Work}\label{sec:applications}

Recent research has produced several approaches to secure edge computing environments through the usage of LLMs. 
For instance, $SecurityBERT$ \cite{ferrag2024revolutionizing} is an approach for IoT/IIoT malware detection through a privacy-focused BERT-based lightweight model.
The system can identify various malware types, including ransomware, spyware, and keyloggers. Its novel privacy-preserving fixed-length encoding (PPFLE) technique converts unstructured network data into BERT-compatible text while using cryptographic hash functions to protect sensitive information.  
Trained on real network traffic, logs, and alerts, $SecurityBERT$ achieves  98.2\% accuracy in distinguishing between 14 malware types.

Zhang et al. (2024) \cite{zhang2024edgeshard} developed $EdgeShard$, a framework that optimizes LLM deployment on edge computing by enabling collaborative inference among heterogeneous edge devices and cloud servers. 
$EdgeShard$'s innovative approach involves breaking down LLMs into smaller components distributed across multiple devices while using optimization algorithms to determine optimal resource allocation and model partitioning. 
Experiments conducted with Llama2 models demonstrate significant improvements in both latency reduction and throughput compared to baseline methods (e.g., strategies to process all data on the edge device or split processing equally between the edge device and the cloud). However, $EdgeShard$ is not used for malware detection, rather for text generation.

Additionally, \cite{chen2024adaptive} presents a novel framework designed to optimize the deployment of LLMs in edge computing settings through model-based reinforcement learning (MBRL). 
The approach focuses on determining the most effective splitting point within the LLM architecture to improve efficiency and performance under wireless network conditions.
By modelling the problem as a Markov Decision Process and integrating a reward surrogate model, the framework adeptly balances inference performance with computational demands on end-user devices.
Experimental results highlight the framework's success in achieving this balance.  However, \cite{chen2024adaptive} has not been designed for malware detection tasks.

Ravi et al. (2023) \cite{ravi2023vit4mal} presents an approach to malware detection in resource-constrained edge devices by utilizing a lightweight vision transformer (ViT) model. 
Their method converts executable bytecodes into images for analysis, implementing the system on a PYNQ Z1 FPGA board with limited resources. The ViT model demonstrated superior accuracy to traditional CNN architectures, like InceptionV3 and ResNet-50, proving particularly effective for real-time edge device applications.

While the above-mentioned proposals present promising results for securing edge computing environments through LLMs, their approaches involve model optimization or modification. 
Our proposal investigates the performance of LLM models as vanilla and after fine-tuning, which involves adjusting the model’s pre-trained parameters without changing the model architecture.

Other approaches utilize LLMs for malware detection, though not specifically targeting edge computing. 
Most of them rely on BERT (see Section \ref{sec:background}) and its variants.
For example, SecureBERT \cite{aghaei2022securebert}, a cybersecurity-specific model based on RoBERTa, is trained on a corpus of 1.1 billion words from various online sources. It automates malware and intrusion detection by interpreting cybersecurity texts, such as cyber threat intelligence (CTI) documentation, and enhances malware detection training through its domain-specific understanding.
MalBERT \cite{rahali2021malbert_android} applies BERT for detecting Android malware via static source code analysis. Its extension, MalBERTv2 \cite{rahali2023malbertv2_android}, refines malware identification using code-aware BERT models, pre-tokenization, and relevance-focused training.
\cite{ullah2022explainable} combines textual and visual features for Android malware detection, using a BERT-based transfer learning approach and a novel malware-to-image transformation for enhanced analysis.
AppPoet \cite{zhao2024apppoet} leverages GPT-4 with multi-view prompt engineering to improve Android malware detection by incorporating detailed feature descriptions.
 \cite{xu2021malbert_windows} emphasizes pre-training's role in Windows malware detection through dynamic analysis, showcasing BERT-based models' potential in this area.

\cite{sanchez2024transfer} presents an approach to malware detection with LLMs by examining the patterns of system calls generated by applications. 
The authors utilized a 1TB dataset of system calls collected from a Raspberry Pi 3-based spectrum sensor called MalwSpecSys. This dataset encompasses benign and malicious behaviours, providing a rich and diverse data source for the training and validation of the proposed models. 
The authors evaluate several LLMs, including BERT, DistilBERT, GPT-2, BigBird, Longformer, and Mistral, each characterized by different context sizes that influence their ability to process long sequences of syscalls.
The performance of these models was assessed using various classification metrics, including accuracy, precision, recall, F1-score, Cohen’s Kappa, and Matthews Correlation Coefficient (MCC). 
The results indicate a trend where models with larger context sizes (i.e., the amount of input data the model can process at once)  generally exhibited superior performance, with BigBird and Longformer achieving the highest scores. 
However, while larger contexts can capture more information, they may lead to overfitting and misclassification of certain behaviours, as evidenced by the performance decline of the Mistral model at a context size of 8192 tokens.
The authors suggest that employing ensemble methods, which combine predictions from multiple models with varying context sizes, could further enhance detection accuracy by leveraging the strengths of each model.
The work demonstrates the potential of LLMs in the context of malware detection through syscall analysis.

\cite{omar2024harnessing} presents a comparative analysis of LLMs, specifically BERT and GPT-2, for the detection of zero-day attacks on IoT devices. 
The adopted methodology involves preprocessing three datasets, ToN-IoT, NSL-KDD, and KDD99, before training and fine-tuning the models.
While all the datasets are popular, NSL-KDD and KDD99 are outdated.
The models are deployed for real-time anomaly detection, with performance evaluated using accuracy, precision, recall, and F1-score metrics.
The analysis results indicate that BERT outperforms GPT-2 across multiple metrics, demonstrating its effectiveness in generalizing to various attack types, but does not provide results for the deployment of such models on the device.

 \cite{panebianco2025guessing} presents a comprehensive analysis of the effectiveness and limitations of LLMs for software automatic vulnerability detection in the context of small and medium-sized enterprises (SMEs) that may not have the resources to engage in professional security assessments.
%The authors review existing literature on datasets and model architectures used for this task, highlighting common datasets like Big-Vul \cite{fan2020ac} and Reveal \cite{chakraborty2021deep} and a variety of prompting techniques employed in recent studies. 
The paper evaluates open-source general-purpose models, such as LLAMA 3.2 8B \cite{dubey2024llama}, and commercials models, such as GPT-4o mini \cite{gptcore}, noting that many of them are subject to overfitting, leading to unreliable performance metrics for the considered use case. 
The findings indicate that while LLMs are showing promising results, their precision in recognizing software vulnerabilities is inadequate.
Additionally, the paper underscores the necessity for evaluation datasets that are disjoint from training data to measure model performance accurately.

%section closing:
Apart from \cite{omar2024harnessing}, the above-mentioned proposals focus on software vulnerabilities, while we focus on malware detection on edge devices.
% Additionally, none of them has considered the deployment feasibility of LLM models on edge devices.
Additionally, none of these proposals produces a comprehensive performance and feasibility analysis of deploying an LLM on an edge device, which is one of the key contributions of our paper.
% Finally, none of these solutions considers multiple datasets and models in the same analysis, focusing on multiple models or datasets simultaneously. 
% Finally, none of these solutions consider focusing on multiple models and datasets simultaneously. 
% Finally, all these solutions focus on multiple models or multiple datasets. We provide a more comprehensive analysis by considering multiple models and datasets.
Finally, none of these solutions provides a comprehensive analysis considering multiple models and datasets simultaneously, as the one we conduct.

%The following sections will present the challenges of performing such an analysis, along with the experimental result of the proposed solution.

\section{Deploying Large Language Models for Malware Detection on Edge Devices}\label{sec:challenges+architecture}

The key idea of our proposal is deploying LLMs on an edge devices to analyze network traffic in real-time and perform malware detection.   We expect two main benefits from this solution. The first is an improved response time and reduced network latency due to the analysis being performed locally on the edge device without sending data to a remote server. The second is that having the model analyzing the local network brings the capability of rapid adaptation and response to evolving threats.   

However, in designing LLM-based malware detection at the edge level, we have taken into account three main requirements that drove the final architecture.

\begin{itemize}
\item \textbf{Feasibility}. Using LLMs for malware detection requires significant computational and memory demands. Such requirements often exceed the capabilities of the majority of edge devices, typically constrained by limited computational power, storage, and energy resources. As such, the naive solution of deploying the model directly on edge devices cannot be applied. This requires {\em investigating alternative architectures/solutions, able to trade off the limited resources demands and the model's accuracy}.

\item \textbf{Accuracy}. Although LLMs excel in natural language understanding and pattern recognition, maintaining consistently high accuracy under resource constraints poses a considerable challenge. 
A possible solution is to specialize the LLM with tailored training through fine-tuning. 
Fine-tuning \cite{howard2018universal} is a transfer learning technique where a pre-trained model is adapted to a specific task by further training it on a smaller, domain-specific dataset.
Pre-trained models, such as those derived from large-scale datasets (e.g., BERT, distilBERT), already possess a generalized understanding of patterns and structures within data. 
Fine-tuning leverages this pre-existing knowledge by focusing on task-specific nuances, enabling the model to achieve better performance with fewer computational resources compared to training from scratch. 
During fine-tuning, the model weights are adjusted based on the new dataset using supervised learning methods. 
The process also involves adapting the model's architecture for the specific task.
This can include adding a classification head or specialized final layers that transform the model's learned representation into the desired output format (e.g., probability scores for classification tasks).
It is also possible to modify the model's architecture more extensively, such as adding pooling layers for text classification or cross-attention mechanisms for question answering. 
The model is then trained by defining an appropriate loss function (e.g., binary cross-entropy loss for binary classification) and optimizing the parameters using algorithms such as $AdamW$ \cite{loshchilov2017decoupled}.
%Additionally to fine-tuning, other alternatives can be exploited to train the model with a small number of labelled examples, such as one-shot or few-shot learning.

\begin{figure*}[!htbp]
\centering
\includegraphics[width=0.75\textwidth, keepaspectratio]{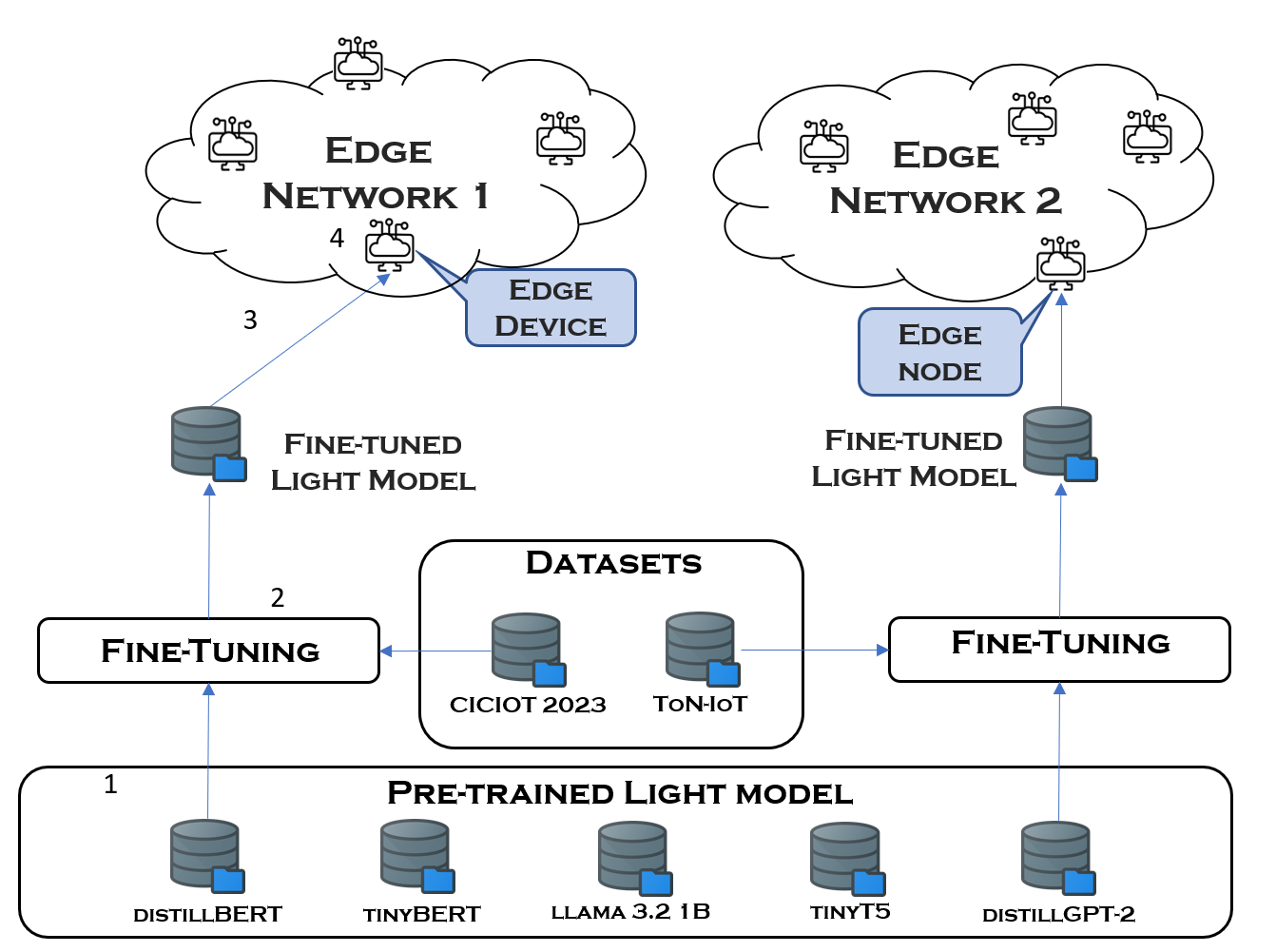}
\caption{Proposed architecture}\label{fig:EdgeLayerArchitecture}
\end{figure*} 

It is expected that a fine-tuned model obtains improved accuracy without increasing device resource consumption.
However, fine-tuning requires significant resource usage in the additional training phases. Thus, in designing an LLM-based malware detection solution for edge devices, a relevant challenge is to {\em decide if fine-tuning is needed and, in that case, how to perform it effectively}.

\item \textbf{Adaptability}. Applying LLMs directly on edge devices brings the benefits of localized threat identification. However, models analyzing a single network flow may lead to poor adaptability to other network flows. This is particularly relevant in a malware detection scenario, where malware patterns and attack behaviors are not uniform across different network environments. This is further exacerbated if fine-tuning is adopted. Indeed, a model that has been fine-tuned on a dataset representing a particular network flow may become very good at detecting threats within that environment but struggle to recognize attacks in other contexts. Moreover, if the dataset used for fine-tuning lacks diversity, the model may overfit to specific traffic patterns or attack signatures rather than learning more generalizable features.
In general, overfitting occurs when a model learns to perform exceptionally well on the training data but fails to generalize to unseen data. 
This happens when the model captures noise, outliers, or overly specific patterns in the training set, mistaking them for general trends. 
Overfitting typically arises from excessive training, insufficiently diverse datasets, or an overly complex model architecture that matches the training data too closely. 
The primary consequence of overfitting is poor performance on new data, making the model less reliable in real-world applications.
In malware detection tasks, an overfitted model might excel at identifying known malware signatures but struggle to detect novel threats, thus undermining its utility.

The risk of overfitting during the fine-tuning process can make LLMs susceptible to biases and inaccuracies, and this could result in biased predictions or an inability to identify novel threats.
It is thus mandatory to {\em experimentally evaluate whether  LLM-based malware detection can adapt to new network flows}, that is, if it is able to transfer knowledge learned from one network flow to another.
To mitigate overfitting, techniques such as k-fold cross-validation, regularization, and the use of diverse training datasets can be employed \cite{santos2022avoiding,ghojogh2019theory}, to ensure the model retains its generalizability and robustness across various scenarios.
\end{itemize}

%% Architettura -
In our proposed architecture, to address the feasibility requirement, rather than LLM models,  we focus on small language models (SLMs) \cite{xu2024survey}. 

Selecting the SLM to be adopted is not an easy task since several factors need to be considered (e.g., available resources, desired level of accuracy). To better understand which variables impact the model selection, we experimentally evaluate several SLMs simulating different device settings. In particular,   we consider the distilled and tiny models reported in Table \ref{tab:LLMsmodels}.   Additionally, as mentioned in Section \ref{sec:background}, related literature \cite{yang2024harnessing, chang2024survey} claims that models from the encoder-only category should be preferred over decoder-only and encoder-decoder, as they are specifically tailored for classification tasks. Therefore, we experimentally evaluate if this holds also for malware detection.
As such, we select at least one representative from each model category (i.e., encoder-only, decoder-only and encoder-decoder), considering DistilGPT-2, DistilBERT, TinyBERT, LLaMA 3.2 1B, and TinyT5 (see Figure \ref{fig:EdgeLayerArchitecture} (1)), to evaluate which is the most suitable for malware detection.

However, even if we leverage on distilled/tiny models, we need to carefully evaluate whether these models can be efficiently deployed on edge devices that are very heterogeneous in computational power and memory size. Thus, we conduct extensive experiments to determine if edge devices with different capabilities, varying from simple devices (e.g., CCTV cameras) to edge nodes (e.g., Jetson Nano), can support efficient SLM-based real-time malware detection.

To properly address the accuracy requirement, we first run a set of experiments to verify if fine-tuning is required. In particular, we evaluate the SLM models' accuracy under zero-shot settings, that is, when fine-tuning is not applied to the model, and under the few-shot learning setting, that is, when only a few dataset items are used to train the model. Initial results (see Section \ref{sec:expresults}) show that while SLMs can still perform basic tasks (e.g., keyword matching), their accuracy is not optimal for detecting complex malware patterns.

Based on these results, we propose a fine-tuned SML-based malware detection approach, graphically depicted in Figure \ref{fig:EdgeLayerArchitecture}. Adopting fine-tuning requires the selection of a suitable dataset for training.
With this respect, we opt to select different training datasets, each one able to tailor the model to specific malware patterns. This allows to have edge devices in the architecture fine-tuned to detect specific types of malware, allowing specialized and targeted threat identification.

Selecting which malware type to associate with each edge device considers the device's context and functionality. For instance, if the edge device manages network traffic, it might be more susceptible to botnet attacks, or if it handles sensitive financial or personal data, it may need enhanced protection against ransomware (see Figure \ref{fig:EdgeLayerArchitecture} (2)).

On one hand, this specialization is expected to improve the overall system's ability to detect a wide range of malware efficiently and enable faster response times to potential security incidents. On the other hand, this could increase the risk of poor adaptability of the model to network flows that differ from the training data (i.e., \textbf{adaptability} requirement).

To evaluate this risk, we rigorously assess the models' ability to adapt to other network flows. In particular, we adopt a {\em cross-dataset validation} strategy, where we train the models on one dataset (e.g., EdgeIIoTset) and test them on a different dataset (e.g., Ton-IoT). Also, to ensure the robustness and generalizability of the fine-tuned models and to minimize the risk of overfitting, we employed the k-fold cross-validation technique. The results of the tests are detailed in Section \ref{sec:expresults}.

\section{Experimental Results}\label{sec:expresults}

In this section, we provide the results of our extensive experimental evaluation, with the aim of verifying whether LLMs can be effectively used for malware detection at the edge.  In particular, the experiments have been design to evaluate the \textbf{accuracy} (see Sections \ref{sub:zero} and \ref{sub:fine}),  \textbf{adatability} (see Section \ref{sub:cross}), and \textbf{feasability} (see Section \ref{sub:device}).
%machine specification

\subsection{Settings}
Our experiments were conducted on two different machines: a high-performance system equipped with an AMD EPYC 7742 64-core Processor, 32GB RAM, and an NVIDIA A100-SXM4-80GB GPU, and a laptop featuring an Intel i7-10710U 6-core CPU (4.7 GHz), 16GB DDR4 RAM, and an NVIDIA GTX 1650 Max-Q 4GB GPU. 
In the following, we do not specify the machine used in each experiment, as it only impacts the training time for fine-tuning, while the other performance indicators remain identical. The only exception is for training large datasets (e.g., CIC IoT 2023), where we had to use the high-performance system, as the laptop lacked sufficient memory to complete the process.

% \usepackage{booktabs}
% \usepackage{multirow}
% \usepackage{graphicx}
\begin{table*}[]
\resizebox{\textwidth}{!}{%
\begin{tabular}{@{}|lcl|@{}}
\toprule
\multicolumn{1}{|c|}{\textbf{Metric}} & \multicolumn{1}{c|}{\textbf{Formula}} & \textbf{Description} \\ \midrule
\multicolumn{3}{|c|}{Model performance metrics} \\ \midrule
\multicolumn{1}{|l|}{\multirow{3}{*}{Accuracy}} & \multicolumn{1}{c|}{\multirow{3}{*}{$\displaystyle \frac{TP + TN}{TP + TN + FP + FN}$}} & \multirow{3}{*}{\begin{tabular}[c]{@{}l@{}}Evaluate the overall correctness of the model by comparing the number of correctly \\ classified instances (both positive and negative) to the total number of instances\end{tabular}} \\
\multicolumn{1}{|l|}{} & \multicolumn{1}{c|}{} &  \\
\multicolumn{1}{|l|}{} & \multicolumn{1}{c|}{} &  \\ \midrule
\multicolumn{1}{|l|}{\multirow{3}{*}{Precision}} & \multicolumn{1}{c|}{\multirow{3}{*}{$\displaystyle \frac{TP}{TP + FP}$}} & \multirow{3}{*}{\begin{tabular}[c]{@{}l@{}}Measures the proportion of true positives among all instances \\ predicted as positive (proportion of estimated positives that are actually positive)\end{tabular}} \\
\multicolumn{1}{|l|}{} & \multicolumn{1}{c|}{} &  \\
\multicolumn{1}{|l|}{} & \multicolumn{1}{c|}{} &  \\ \midrule
\multicolumn{1}{|l|}{\multirow{3}{*}{Recall}} & \multicolumn{1}{c|}{\multirow{3}{*}{$\displaystyle \frac{TP}{TP + FN}$}} & \multirow{3}{*}{Proportion of real positives that are estimated positive, known as True Positive Rate} \\
\multicolumn{1}{|l|}{} & \multicolumn{1}{c|}{} &  \\
\multicolumn{1}{|l|}{} & \multicolumn{1}{c|}{} &  \\ \midrule
\multicolumn{1}{|l|}{\multirow{3}{*}{F1-score}} & \multicolumn{1}{c|}{\multirow{3}{*}{$\displaystyle 2 \times \frac{\text{Precision} \times \text{Recall}}{\text{Precision} + \text{Recall}}$}} & \multirow{3}{*}{\begin{tabular}[c]{@{}l@{}}Is the harmonic mean of precision and recall, \\ balancing both false positives and false negatives\end{tabular}} \\
\multicolumn{1}{|l|}{} & \multicolumn{1}{c|}{} &  \\
\multicolumn{1}{|l|}{} & \multicolumn{1}{c|}{} &  \\ \midrule
\multicolumn{3}{|c|}{Models performance on device} \\ \midrule
\multicolumn{1}{|l|}{Memory for input tensors} & \multicolumn{1}{c|}{$\displaystyle \text{Batch size} \times \text{S\_length} \times \text{FPA}$} & \begin{tabular}[c]{@{}l@{}}Memory consumed by the input data (tokens or sequences) \\ before being processed by the model\end{tabular} \\ \midrule
\multicolumn{1}{|l|}{Memory for activations} & \multicolumn{1}{c|}{$\displaystyle \text{Batch size} \times \text{S\_length} \times \text{Hidden size} \times \text{N\_layers} \times \text{FPA}$} & \begin{tabular}[c]{@{}l@{}}Memory required for storing intermediate activations (hidden states) \\ during forward and backward passes\end{tabular} \\ \midrule
\multicolumn{1}{|l|}{Memory for output tensors} & \multicolumn{1}{c|}{$\displaystyle \text{Batch size} \times \text{S\_length} \times \text{Vocab. size} \times \text{FPA}$} & Memory allocated for the output of the model \\ \midrule
\multicolumn{1}{|l|}{\multirow{3}{*}{Latency (prediction time)}} & \multicolumn{1}{c|}{\multirow{3}{*}{$\displaystyle \frac{\text{Model total FLOPs}}{\text{Hardware FLOPs}}$}} & \multirow{3}{*}{\begin{tabular}[c]{@{}l@{}}Time required for the model to generate predictions on given input data,\\ based on the available computational resources\end{tabular}} \\
\multicolumn{1}{|l|}{} & \multicolumn{1}{c|}{} &  \\
\multicolumn{1}{|l|}{} & \multicolumn{1}{c|}{} &  \\ \midrule
\multicolumn{1}{|l|}{FLOPS(head)} & \multicolumn{1}{c|}{$ \displaystyle 4 \times (\text{S\_length})^2 \times \text{Hidden size} \times \text{N\_heads}$} & Computational complexity of the self-attention mechanism in the transformer model \\ \midrule
\multicolumn{1}{|l|}{FLOPS(feedforward)} & \multicolumn{1}{c|}{$\displaystyle 8 \times \text{S\_length} \times (\text{Hidden size})^2$} & \begin{tabular}[c]{@{}l@{}}Computational cost of the feedforward layers, \\ which transform the embeddings within each transformer block\end{tabular} \\ \midrule
\multicolumn{1}{|l|}{Total FLOPS} & \multicolumn{1}{c|}{$\displaystyle \text{ N\_layers} \times (\text{FLOPs (head)} + \text{FLOPs (feedforward)})$} & Total number of FLOPs required to process a single forward pass through the model \\ \bottomrule
\end{tabular}%
}
\vspace{+0.1em}
\caption{Performance metrics used in the evaluation. Total FLOPs (floating-point operations) refer to the number of arithmetic operations involving floating-point numbers that the model performs. N\_head is the number of attention heads, FPA denotes floating-point accuracy, S\_length is the sequence length.}
\label{tab:metrics}
\end{table*}

Experiments were conducted starting from multiple pre-trained language models. Each pre-trained model was retrieved for the Pytorch framework from the official repository on Hugginfaces\footnote{https://huggingface.co/}, which can be considered a de facto standard for the open-source code of NLP and machine learning models.
The models considered in our experiments  include: 
DistilGPT2,\footnote{https://huggingface.co/distilbert/distilgpt2} 
DistilBERT,\footnote{https://huggingface.co/distilbert/distilbert-base-uncased}
TinyBERT,\footnote{https://huggingface.co/huawei-noah/TinyBERT\_General\_4L\_312D} 
LLama 3.2 1B\footnote{https://huggingface.co/meta-llama/Llama-3.2-1B-Instruct}, 
and TinyT5.\footnote{https://huggingface.co/t5-efficient-tiny}
The models were trained on the four selected datasets, namely  X-IIoTID, CIC IoT 2023, Edge-IIoTset, and TON-IoT,  described in Section \ref{sec:background}.

%finetuning process details
Datasets undergo a preparation phase that involves loading data from CSV files, encoding target labels using Scikit-learn's LabelEncoder, creating multilabel targets by aggregating encoded label columns, and concatenating feature columns for model input. 
The resulting dataset was split into training (60\%) and test (40\%) subsets.
Model initialization utilized pre-trained models from the Hugging Face repository, configuring classification heads according to the number of distinct classes (e.g., $GPT2ForSequenceClassification$ for distillGPT-2).
Models and tokenizers are then loaded onto the target computing devices.
Training configuration employed the $AdamW$ optimizer with a learning rate of $2 \times 10^{-5}$ and $BCEWithLogitsLoss$ as the loss function, combining sigmoid activation with binary cross-entropy. 
The training procedure is executed on at least four epochs, with each epoch comprising forward propagation, loss computation, back-propagation, and optimization steps.
Validation occurred after each epoch to monitor performance metrics.

\begin{table}[!htbp]
\centering
\resizebox{\columnwidth}{!}{%
\begin{tabular}{@{}|l|l|c|c|c|c|@{}}
\toprule
\multicolumn{1}{|c|}{Model}     & \multicolumn{1}{c|}{Dataset} & Accuracy & Precision & Recall & F1-Score \\ \midrule
{\multirow{4}{*}{distilGPT2}}   & X-IIoTID                      & 0.4950  & 0.2676    & 0.4950 & 0.3460 \\ \cmidrule(l){2-6} 
                                & Edge-IIoTset                  & 0.7164  & 0.5020    & 0.7164 & 0.6164 \\ \cmidrule(l){2-6} 
                                & TON-IoT                       & 0.7189  & 0.5124    & 0.7189 & 0.5985 \\ \cmidrule(l){2-6} 
                                & CIC IoT 2023                  & 0.7290  & 0.5230    & 0.7290 & 0.6089 \\ \midrule
\multirow{4}{*}{distilBERT}     & X-IIoTID                      & 0.5456  & 0.2427    & 0.5135 & 0.3484 \\ \cmidrule(l){2-6} 
                                & Edge-IIoTset                  & 0.7193  & 0.5305    & 0.7284 & 0.6025 \\ \cmidrule(l){2-6} 
                                & TON-IoT                       & 0.7189  & 0.5124    & 0.7189 & 0.5985 \\ \cmidrule(l){2-6} 
                                & CIC IoT 2023                  & 0.7290  & 0.5230    & 0.7290 & 0.6089 \\ \midrule
\multirow{4}{*}{TinyBERT}       & X-IIoTID                      & 0.5135  & 0.2637    & 0.5062 & 0.3514 \\ \cmidrule(l){2-6} 
                                & Edge-IIoTset                  & 0.7284  & 0.5305    & 0.7284 & 0.6139 \\ \cmidrule(l){2-6} 
                                & TON-IoT                       & 0.7230  & 0.5195    & 0.7230 & 0.6048 \\ \cmidrule(l){2-6} 
                                & CIC IoT 2023                  & 0.7344  & 0.5289    & 0.7344 & 0.6152 \\ \midrule
\multirow{4}{*}{Llama 3.2 1B}   & X-IIoTID                      & 0.5070  & 0.2626    & 0.5070 & 0.3460 \\ \cmidrule(l){2-6} 
                                & Edge-IIoTset                  & 0.5833  & 0.5981    & 0.5833 & 0.5906 \\ \cmidrule(l){2-6} 
                                & TON-IoT                       & 0.4522  & 0.2456    & 0.7009 & 0.3607 \\ \cmidrule(l){2-6} 
                                & CIC IoT 2023                  & 0.4410  & 0.2378    & 0.6947 & 0.3624 \\ \midrule
\multirow{4}{*}{TinyT5}         & X-IIoTID                      & 0.5135  & 0.2425    & 0.5135 & 0.3484 \\ \cmidrule(l){2-6} 
                                & Edge-IIoTset                  & 0.7164  & 0.5346    & 0.7274 & 0.6223 \\ \cmidrule(l){2-6} 
                                & TON-IoT                       & 0.7180  & 0.5158    & 0.7180 & 0.6005 \\ \cmidrule(l){2-6} 
                                & CIC IoT 2023                  & 0.7208  & 0.5175    & 0.7208 & 0.6028 \\ \bottomrule
\end{tabular}
}
\vspace{+0.1em}
\caption{Results for zero-shot tests with binary classification}
\label{tab:zeroshot}
\end{table}

\begin{table*}[]
\centering
\resizebox{\textwidth}{!}{%
\begin{tabular}{@{}|l|l|c|c|c|c|c|c|c|c|c|@{}}
\toprule
\multicolumn{1}{|c|}{Model} &
  \multicolumn{1}{c|}{Dataset} &
  \multicolumn{1}{c|}{Epochs} &
  \multicolumn{1}{c|}{Train. Time} &
  \multicolumn{1}{c|}{Train.Loss} &
  \multicolumn{1}{c|}{Train. Accuracy} &
  \multicolumn{1}{c|}{Test Loss} &
  \multicolumn{1}{c|}{Test Accuracy} &
  \multicolumn{1}{c|}{Precision} &
  \multicolumn{1}{c|}{Recall} &
  \multicolumn{1}{c|}{F1-score}
  \\ \midrule
{\multirow{4}{*}{distilGPT2}}  & X-IIoTID     &  4 &  2 &  0.0730 &  97.57\% &  0.0561 &  98.05\% &  0.9790 & 0.9760 & 0.9775 \\ \cmidrule(l){2-11} 
                                & Edge-IIoTset & 4 & 2   & 0.0391 & 99.12\% & 0.0257 & 99.22\% & 0.9920 & 0.9900 & 0.9910 \\ \cmidrule(l){2-11} 
                                & TON-IoT      & 4 & 2   & 0.0788 & 98.55\% & 0.0543 & 98.57\% & 0.9855 & 0.9835 & 0.9845 \\ \cmidrule(l){2-11} 
                                & CIC IoT 2023 & 4 & 2   & 0.0866 & 97.45\% & 0.0754 & 97.55\% & 0.9740 & 0.9720 & 0.9737 \\ \midrule
\multirow{4}{*}{distilBERT}    & X-IIoTID     & 4 & 2   & 0.0572 & 99.20\% & 0.0477 & 99.14\% & 0.9910 & 0.9912 & 0.9911 \\ \cmidrule(l){2-11} 
                                & Edge-IIoTset & 4 & 2   & 0.0500 & 99.07\% & 0.0376 & 99.21\% & 0.9912 & 0.9900 & 0.9906 \\ \cmidrule(l){2-11} 
                                & TON-IoT      & 4 & 2   & 0.0688 & 98.55\% & 0.0529 & 98.40\% & 0.9835 & 0.9830 & 0.9836 \\ \cmidrule(l){2-11} 
                                & CIC IoT 2023 & 4 & 2   & 0.1048 & 96.73\% & 0.0925 & 96.88\% & 0.9555 & 0.9850 & 0.9852\\ \midrule
\multirow{4}{*}{TinyBERT}       & X-IIoTID     & 4 & 0.5 & 0.3075 & 86.70\% & 0.2988 & 87.30\% & 0.8450 & 0.8350 & 0.8390\\ \cmidrule(l){2-11} 
                                & Edge-IIoTset & 4 & 0.5 & 0.1927 & 93.16\% & 0.1823 & 94.07\% & 0.9400 & 0.9300 & 0.9350 \\ \cmidrule(l){2-11} 
                                & TON-IoT      & 4 & 0.5 & 0.3086 & 88.42\% & 0.2843 & 88.92\% & 0.8890 & 0.8780 & 0.8815 \\ \cmidrule(l){2-11} 
                                & CIC IoT 2023 & 4 & 0.5 & 0.1617 & 96.30\% & 0.1534 & 96.30\% & 0.9615 & 0.9600 & 0.9607 \\ \midrule
\multirow{4}{*}{Llama 3.2 1B}   & X-IIoTID     & 4 & 3   & 0.3060 & 89.16\% & 0.2917 & 87.40\% & 0.9702 & 0.9689 & 0.9640 \\ \cmidrule(l){2-11} 
                                & Edge-IIoTset & 4 & 3   & 0.1716 & 94.17\% & 0.1701 & 93.88\% & 0.9913 & 0.9885 & 0.9755\\ \cmidrule(l){2-11} 
                                & TON-IoT      & 4 & 3   & 0.2988 & 91.98\% & 0.2980 & 90.01\% & 0.9556 & 0.9508 & 0.9462\\ \cmidrule(l){2-11} 
                                & CIC IoT 2023 & 4 & 3   & 0.1514 & 96.01\% & 0.1412 & 96.25\% & 0.9667 & 0.9675 & 0.9681\\ \midrule
\multirow{4}{*}{TinyT5}      & X-IIoTID     & 50 & 8.3 & 0.0635 & 96.67\% & 0.0501 & 97.00\% & 0.9703 & 0.9680 & 0.9692 \\ \cmidrule(l){2-11} 
                                & Edge-IIoTset & 15 & 2.5 & 0.1409 & 97.50\% & 0.0269 & 97.12\% & 0.9757 & 0.9740 & 0.9748 \\ \cmidrule(l){2-11} 
                                & TON-IoT      & 15 & 2.5 & 0.1292 & 95.85\% & 0.0554 & 95.76\% & 0.9369 & 0.9350 & 0.9359 \\ \cmidrule(l){2-11} 
                                & CIC IoT 2023 & 100 & 166 & 0.1834 & 90.00\% & 0.2421 & 91.25\% & 0.9125 & 0.9100 & 0.9112\\ \bottomrule
\end{tabular}
}
\vspace{+0.1em}
\caption{Results for the few-shot test. Training time is minutes}
\label{tab:fewshot}
\end{table*}

Table \ref{tab:metrics} reports all the performance metrics used in our experiments.
The model's performance is evaluated using the following metrics: accuracy, precision, recall and F1-score.
In their definitions, TP (true positives) denotes data items  (i.e., network flow) correctly labeled as positive, that is, with malware; FN (false negatives) denotes the data items incorrectly labeled as negative when they are actually positive;  FP (false positives) denotes those labeled as positive when they are actually negative; whereas TN (true negatives) denotes the number of those correctly labeled as negative, i.e., without malware.
Moreover, in the table, batch size refers to the number of samples processed simultaneously during training, floating-point accuracy (FPA) indicates the precision level of numerical computations, whereas sequence length (S\_lengthS) defines the maximum number of tokens the model can process in a single forward pass.
The literature provides formulas that can be exploited to calculate the model performances on the device theoretically \cite{vaswani2017attention,wolf2020transformers,EfficentTrasformerTAY}.

%references
% "Efficient Transformer: A Survey" (Tay et al., 2020), Section 2.2: Complexity Analysis
% "Attention Is All You Need" (Vaswani et al., 2017), Table 1: Computational Complexity per Layer
% "Transformers: State-of-the-Art Natural Language Processing" (Wolf et al., 2020)

%overfitting techniques
To mitigate the risk of overfitting during model training, we apply 
k-fold cross-validation. 
Specifically, the process was conducted with $k=5$, ensuring a robust evaluation by splitting the dataset into five distinct folds, each serving as a validation set. In contrast, the remaining folds were used for training.
The application of this technique demonstrated consistency in the results, as the performance metrics obtained with k-fold cross-validation were identical to those achieved without its application.
This outcome suggests that overfitting was effectively prevented, ensuring the model's ability to generalize well to unseen data. 
Incorporating this approach further validated the training process, reinforcing the model's reliability for deployment in real-world scenarios.

%feature extraction
Moreover, we applied feature extraction techniques to identify whether relevant attributes within the dataset were present. 
Those techniques included lasso, recursive feature elimination (RFE), principal component analysis (PCA), random forest and correlation matrix.   
For instance, when applying the lasso technique to the X-IIoTID dataset, the analysis identified only 6 out of 65 features as significant for the task. 
Instead, the correlation matrix and Random Forest methods removed 20 features, whereas RFE and PCA retained 25.
% This indicates that most features may not contribute meaningfully to the model's performance.
However, to ensure a comprehensive evaluation, the test sets were run using the full datasets without applying any feature pruning. 

%eveluation sets & machine specification
To cope with the challenges presented in Section \ref{sec:challenges+architecture}, we design and perform three sets of experiments that are described in what follows.

% ###################  ZERO SHOT TEST#########################
\subsection{Zero-shot classification}\label{sub:zero}
In the first set of experiments, we evaluate the model's performance using a zero-shot classification approach. 
Zero-shot classification is a scenario where the model is applied to a task without any task-specific fine-tuning.
The model is expected to generalize from its pre-trained knowledge to perform the classification task.
In our context, the task is to classify whether the data represents benign or malicious behaviour. % they were provided with tuples from the dataset.
The outcomes of this evaluation are summarized in Table \ref{tab:zeroshot}, which reports accuracy, precision, recall, and F1-score for each considered model and dataset.

%Discussion zero shot.
The results in Table \ref{tab:zeroshot} indicate that the models achieve a maximum accuracy of 72\%, leading to suboptimal performance.
Among the models, we see how encoder-only models, namely distilBERT and distilGPT2, perform slightly better than the others.
This aligns with our prediction and results reported in related literature (e.g., \cite{panebianco2025guessing}) and demonstrates that considered LLM models are not ready to be used without fine-tuning for the malware detection task.

% ###################  FINETUNING:  FEW SHOT e FULL FINETUNING TEST#########################
\subsection{Fine-tuning classification}\label{sub:fine}
In the second set of experiments, we fine-tune the models and measure their accuracy under two distinct scenarios. 

The first scenario, referred to as the same-dataset approach, involves training the model on a specific dataset (D1) and evaluating its accuracy on the same dataset. 
We evaluate the accuracy of this approach by exploiting two approaches for the dataset samples, one named few-shot and the other complete.
The few-shot approach involves training the model with a small number of labelled examples.
In our experiments, we use 30000 lines of the datasets ($\sim$30 MB) for the few-shot. 
The key difference is that in the few-shot approach, the model adapts to the task with limited samples, unlike zero-shot, where no task-specific data is provided.
The test we run uses a binary classification approach.
Instead, for the fine-tuning case, the model undergoes more extensive training using a significantly larger set of labelled data. 
In our case, we exploit the full size of the target datasets.

% same dataset - few shots e full fine-tune discussion
We show the results for the few-shot scenario in Table \ref{tab:fewshot}, whereas Table \ref{tab:finetune} reports the result for the complete case.  
The table lists the number of epochs needed to achieve relevant accuracy and the time needed to complete the training phase.
At the end of each epoch, the training loss and accuracy are calculated; we list only the result of the last epoch. 
The test results are calculated after the training phase, including precision, recall, and f1-score.
We recall that loss expresses how far the model's predictions are from the actual labels, quantifying the error.
A lower loss and higher accuracy indicate better performance.
%accuracy è già spiegata nella tab delle metriche

The fine-tuning results show a significantly increased accuracy moving from the few-shot to the complete approach.
The results indicate that all the considered models achieve an accuracy on average of 94\%  with few shots and 99\% for complete.
However, we note how TinyBERT achieves the best performance, taking into account the training time and number of epochs.
The reason for the worst performance of TinyT5 could be related to its lesser number of heads. 
Although such a result may indicate overfitting, we can exclude it as the difference between train and test loss is minimal, and the results from the k-fold cross-validation are aligned.

\begin{table*}[!htbp]
\centering
\resizebox{\textwidth}{!}{%
\begin{tabular}{@{}|l|l|c|c|c|c|c|c|c|c|c|@{}}
\toprule
\multicolumn{1}{|c|}{Model} &
  \multicolumn{1}{c|}{Dataset} &
  \multicolumn{1}{c|}{Epochs} &
  \multicolumn{1}{c|}{Train Time} &
  \multicolumn{1}{c|}{Train Loss} &
  \multicolumn{1}{c|}{Train Accuracy} &
  \multicolumn{1}{c|}{Test Loss} &
  \multicolumn{1}{c|}{Test Accuracy} &
  \multicolumn{1}{c|}{Precision} &
  \multicolumn{1}{c|}{Recall} &
  \multicolumn{1}{c|}{F1-score}
  \\ \midrule
\multirow{4}{*}{distilGPT2}   & X-IIoTID     & 4  & 1   & 0.0005 & 99.97\% & 0.0010 & 98.05\% & 0.9980 & 0.9805 & 0.9891 \\ \cmidrule(l){2-11} 
                              & Edge-IIoTset & 4  & 4   & 0.0012 & 99.97\% & 0.0002 & 99.97\% & 0.9990 & 0.9997 & 0.9993 \\ \cmidrule(l){2-11} 
                              & TON-IoT      & 4  & 24  & 0.0009 & 99.98\% & 0.0007 & 99.98\% & 0.9991 & 0.9998 & 0.9994 \\ \cmidrule(l){2-11} 
                              & CIC IoT 2023 & 4  & 80  & 0.0008 & 99.96\% & 0.0009 & 99.96\% & 0.9989 & 0.9996 & 0.9992 \\ \midrule
\multirow{4}{*}{distilBERT}   & X-IIoTID     & 4  &  1  & 0.0007 & 99.95\% & 0.0012 & 98.98\% & 0.9985 & 0.9898 & 0.9941 \\ \cmidrule(l){2-11} 
                              & Edge-IIoTset & 4  &  4  & 0.0015 & 99.94\% & 0.0005 & 99.95\% & 0.9988 & 0.9995 & 0.9991 \\ \cmidrule(l){2-11} 
                              & TON-IoT      & 4  & 24  & 0.0011 & 99.96\% & 0.0009 & 99.97\% & 0.9990 & 0.9997 & 0.9993 \\ \cmidrule(l){2-11} 
                              & CIC IoT 2023 & 4  & 80  & 0.0010 & 99.95\% & 0.0011 & 99.94\% & 0.9987 & 0.9994 & 0.9990 \\ \midrule
\multirow{4}{*}{TinyBERT}     & X-IIoTID     & 4  &0.8  & 0.0009 & 99.93\% & 0.0015 & 98.85\% & 0.9982 & 0.9885 & 0.9933 \\ \cmidrule(l){2-11} 
                              & Edge-IIoTset & 4  & 2   & 0.0018 & 99.92\% & 0.0008 & 99.93\% & 0.9985 & 0.9993 & 0.9989 \\ \cmidrule(l){2-11} 
                              & TON-IoT      & 4  & 10  & 0.0013 & 99.94\% & 0.0011 & 99.95\% & 0.9988 & 0.9995 & 0.9991 \\ \cmidrule(l){2-11} 
                              & CIC IoT 2023 & 4  & 40  & 0.0012 & 99.93\% & 0.0013 & 99.92\% & 0.9985 & 0.9992 & 0.9988 \\ \midrule
\multirow{4}{*}{Llama 3.2 1B} & X-IIoTID     & 4  & 1   & 0.0011 & 99.91\% & 0.0018 & 98.75\% & 0.9980 & 0.9875 & 0.9927 \\ \cmidrule(l){2-11} 
                              & Edge-IIoTset & 4  & 4   & 0.0020 & 99.90\% & 0.0010 & 99.91\% & 0.9983 & 0.9991 & 0.9987 \\ \cmidrule(l){2-11} 
                              & TON-IoT      & 4  & 20  & 0.0015 & 99.92\% & 0.0013 & 99.93\% & 0.9986 & 0.9993 & 0.9989 \\ \cmidrule(l){2-11} 
                              & CIC IoT 2023 & 4  & 60  & 0.0014 & 99.91\% & 0.0015 & 99.90\% & 0.9983 & 0.9990 & 0.9986 \\ \midrule
\multirow{4}{*}{TinyT5}       & X-IIoTID     & 25 &  5   & 0.1525 & 98.50\% & 0.2125 & 98.75\% & 0.9275 & 0.9275 & 0.9275 \\ \cmidrule(l){2-11} 
                              & Edge-IIoTset & 8  &  3.2   & 0.1650 & 98.75\% & 0.2250 & 99.00\% & 0.9200 & 0.9200 & 0.9200 \\ \cmidrule(l){2-11} 
                              & TON-IoT      & 8  &  19  & 0.1750 & 98.25\% & 0.2350 & 98.50\% & 0.9150 & 0.9150 & 0.9150 \\ \cmidrule(l){2-11} 
                              & CIC IoT 2023 & 40 & 152 & 0.1834 & 99.00\% & 0.2421 & 99.25\% & 0.9125 & 0.9125 & 0.9125 \\ \bottomrule
\end{tabular}
}
\vspace{+0.1em}
\caption{Results for the finetuning test set. Training time is hours}
\label{tab:finetune}
\end{table*}

% {\multirow{4}{*}{distillGPT2}}  & X-IIoTID     & 4 & 1    & 0.0005 & 99.97\% & 0.0010 & 98.05\% & 0.9980 & & \\ \cmidrule(l){2-11} 
%                                 & Edge-IIoTset & 4 & 4    & 0.0012 & 99.97\% & 0.0002 & 99.97\% & 0.9990 & & \\ \cmidrule(l){2-11} 
%                                 & TON-IoT      & 4 & 80   & 0.0009 & 99.98\% & 0.0007 & 99.98\% & 0.9991 & & \\ \cmidrule(l){2-11} 

\subsection{Cross-dataset validation}\label{sub:cross}
To evaluate the \textbf{adatability} of the proposed architecture, we run an experiment named cross-dataset validation, where the model is fine-tuned on one dataset (D1) but evaluated on a completely different dataset (D2). 
This enables us to evaluate the model's generalisation ability to unseen data.
We show the fine-tuning results in Table \ref{tab:crossVal} for both the few-shot and complete approach.
The table reports the fine-tune model accuracy and shows that the complete approach performs better than the few-shot.
Even if the models mainly achieve satisfactory performance, it is not universally true.
Performance remains suboptimal for some combinations, such as CIC IoT 2023 and X-IIoTID.
The reasons can be attributed to differences in feature distributions in those datasets.
This problem is still an open issue in the literature (see domain shift \cite{kouw2018introduction}) and needs further evaluation.

\begin{table*}[!htbp]
\centering
\resizebox{\textwidth}{!}{%
\begin{tabular}{@{}|l|l|c|c|c|c|c|c|c|c|c|c|c|c|c|@{}}
\toprule

\multirow{2}{*}{Model} &
\multirow{2}{*}{Dataset} & 
\multicolumn{4}{c|}{Few-shot} & 
\multicolumn{4}{c|}{Complete } \\ \cmidrule{3-10} 
 &  & CICIoT2023 & TON-IoT &  Edge-IIoTset & X-IIoTID & CICIoT2023 & TON-IoT &  Edge-IIoTset & X-IIoTID \\ \midrule
distilGPT2 & X-IIoTID     & 15.56\% & 48.54\% & 96.39\% & ---   & 16.28\% & 49.76\%& 97.10\% & ---   \\ \midrule
            & Edge-IIoTset & 15.56\% & 71.35\% & ---     & 68.44\% & 16.68\% & 72.42\% & --- & 69.88\% \\ \midrule
            & TON-IoT      & 15.44\%  & ---     & 61.97\% & 68.44\% & 17.10\% & --- & 62.34\% & 69.78\% \\ \midrule
            & CIC IoT 2023 & ---     & 52.05\% & 93.44\% & 68.44\% & --- & 53.78\% & 94.53\% & 70.14\% \\ \midrule
distilBERT & X-IIoTID     & 15.56\% & 57.31\% & 96.39\% & ---    & 16.88\% & 58.35\% & 97.20\% & --- \\ \midrule
            & Edge-IIoTset & 15.30\%  & 24.56\% & ---   & 19.63\% & 16.82\% & 25.78\% & --- & 20.91\%   \\ \midrule
            & TON-IoT      & 15.56\% & ---     & 96.39\% & 68.44\% & 17.14\% & --- & 97.12\% & 70.35\% \\ \midrule
            & CIC IoT 2023 & ---     & 55.56\% & 96.39\% & 68.44\% & --- & 56.87\% & 97.12\% & 69.25\% \\ \midrule
TinyBERT    & X-IIoTID     & 15.56\% & 55.56\% & 96.39\% & ---     & 16.84\% & 56.88\% & 97.20\% & --- \\ \midrule
            & Edge-IIoTset & 15.56\% & 55.56\% & --- & 68.44\%     & 17.25\%& 57.20\%& --- & 70.28\% \\ \midrule
            & TON-IoT      & 57.78\% & ---     & 14.43\% & 68.44\% & 59.18\% & --- & 15.94\% & 69.78\% \\ \midrule
            & CIC IoT 2023 & ---     & 19.88\% & 12.30\% & 24.25\% & --- & 21.26\% & 14.16\% & 25.68\% \\ \midrule
Llama 3.2 1B& X-IIoTID     & 15.84\% & 50.48\% & 96.44\% & ---  & 17.12\%& 51.68\% & 96.96\% & --- \\ \midrule
            & Edge-IIoTset & 15.28\% & 52.18\% & --- & 65.32\% & 16.72\% & 54.24\% & --- & 66.45\%\\ \midrule
            & TON-IoT      & 15.56\% & --- & 68.46\% & 68.28\% & 16.84\% & --- & 69.78\% & 69.94\%\\ \midrule
            & CIC IoT 2023 & --- & 53.72\% & 94.48\% & 67.99\% & --- & 54.88\% & 95.69\% & 68.74\% \\ \midrule
TinyT5   & X-IIoTID     & 84.44\% & 19.88\% & 12.30\% & --- & 85.80\% & 21.24\% & 13.98\% & --- \\ \midrule
            & Edge-IIoTset & 84.44\% & 19.88\% & --- & 24.25\% & 85.58\% & 21.20\% & --- & 25.78\% \\ \midrule
            & TON-IoT      & 10.01\% & --- & 10.01\% & 28.57\% & 11.94\% & --- & 11.85\% & 30.14\% \\ \midrule
            & CIC IoT 2023 & --- & 55.56\% & 41.31\% & 68.44\% & --- & 56.58\% & 43.16\% & 70.24\% \\ \bottomrule
\end{tabular}
}
\vspace{+0.1em}
\caption{Fine-tuning results for cross-validation with few-shot and complete approach.}
\label{tab:crossVal}
\end{table*}

%overall discussion on finetuning 
Overall, experiments show the importance of fine-tuning, as the results differences between the zero-shot approach and the fine-tuning are significant for all the models.
Also, training the model on a huge dataset is not fully necessary to obtain good results, as we can see from comparing the results between the few-shot and complete approach.
Moreover, effective fine-tuning does not necessarily require access to extensive computational resources.
Even accessible hardware setups can achieve robust performance when sufficient training iterations are conducted.
For instance, the few-shot results were obtained using both of the previously mentioned hardware setups, achieving close results.

% ###################DEVICE TEST SET#########################
\subsection{On device performance}\label{sub:device}
The last set of experiments evaluates the \textbf{feasibility} of our approach on existing edge devices in terms of resource usage and classification time.
%Device virtualization and deployment
For the tests, we select  Raspberry Pi 3 and Jetson Nano devices, two widely used devices for resource-constrained environments.
The first features a CPU with four cores (Cortex-A53, ARMv8, 64-bit) running at 1.4GHz and 1GB of LPDDR2 SDRAM. 
In contrast, Jetson Nano has a quad-core ARM Cortex-A57 CPU, a 128-core Maxwell GPU, and 2GB of LPDDR4 RAM.
To setup the test environment, we deploy the fine-tuned models on a virtual edge device environment. 
We exploit Balena,\footnote{https://www.balena.io/} which offers base images tailored to various edge devices.\footnote{https://docs.balena.io/reference/base-images/base-images-ref/} 
The corresponding device images are therefore retrieved from Balena's repository and deployed into a Docker container. 
The container is configured to emulate the specifications of the selected edge device, including its memory, processing power, and architecture.
Also, the necessary dependencies for model execution are installed, including libraries for managing model inference, such as the Hugging Face Transformers library, and hardware-specific optimizations like NVIDIA TensorRT for the Jetson Nano.
The tests include classification time, resource usage (i.e., CPU, GPU, and memory), and model size. 
To perform the tests, we deploy an application to simulate incoming traffic to the edge device, sending a packet every second. 
On the device, we execute another application consisting of two Python threads.
One thread is entitled to model classification tasks, whereas the other handles the device’s primary function, such as processing CCTV camera streams or monitoring temperature sensors.

\begin{table*}[!htbp]
\resizebox{\textwidth}{!}{%
\begin{tabular}{@{}|llcc|ccc|cc|@{}}
\toprule
\multicolumn{4}{|l|}{}                                                                                                              & \multicolumn{3}{c|}{Jetson nano}                                                                   & \multicolumn{2}{c|}{Raspbeery Pi 3}                 \\ \midrule
\multicolumn{1}{|l|}{Model}                         & \multicolumn{1}{l|}{Dataset}      & \multicolumn{1}{c|}{Size (MB)} & RAM (MB) & \multicolumn{1}{c|}{CPU Usage}  & \multicolumn{1}{c|}{Class. Time CPU (ms)} & Class. Time GPU (ms) & \multicolumn{1}{c|}{Class. Time (sec)} & CPU Usage  \\ \midrule
\multicolumn{1}{|l|}{\multirow{4}{*}{distilGPT2}}  & \multicolumn{1}{l|}{X-IIoTID}     & \multicolumn{1}{c|}{320}       & 750      & \multicolumn{1}{c|}{$\sim$10\%} & \multicolumn{1}{c|}{$\sim$750}            & $\sim$150            & \multicolumn{1}{c|}{$\sim$24}          & $\sim$70\% \\ \cmidrule(l){2-9} 
\multicolumn{1}{|l|}{}                              & \multicolumn{1}{l|}{Edge-IIoTset} & \multicolumn{1}{c|}{320}       & 750      & \multicolumn{1}{c|}{$\sim$10\%} & \multicolumn{1}{c|}{$\sim$750}            & $\sim$150            & \multicolumn{1}{c|}{$\sim$24}          & $\sim$70\% \\ \cmidrule(l){2-9} 
\multicolumn{1}{|l|}{}                              & \multicolumn{1}{l|}{TON-IoT}      & \multicolumn{1}{c|}{320}       & 750      & \multicolumn{1}{c|}{$\sim$10\%} & \multicolumn{1}{c|}{$\sim$750}            & $\sim$150            & \multicolumn{1}{c|}{$\sim$24}          & $\sim$70\% \\ \cmidrule(l){2-9} 
\multicolumn{1}{|l|}{}                              & \multicolumn{1}{l|}{CIC IoT 2023} & \multicolumn{1}{c|}{320}       & 750      & \multicolumn{1}{c|}{$\sim$10\%} & \multicolumn{1}{c|}{$\sim$750}            & $\sim$150            & \multicolumn{1}{c|}{$\sim$24}          & $\sim$70\% \\ \midrule
\multicolumn{1}{|l|}{\multirow{4}{*}{distilBERT}}  & \multicolumn{1}{l|}{X-IIoTID}     & \multicolumn{1}{c|}{260}       & 600      & \multicolumn{1}{c|}{$\sim$10\%} & \multicolumn{1}{c|}{$\sim$750}            & $\sim$150            & \multicolumn{1}{c|}{$\sim$24}          & $\sim$70\% \\ \cmidrule(l){2-9} 
\multicolumn{1}{|l|}{}                              & \multicolumn{1}{l|}{Edge-IIoTset} & \multicolumn{1}{c|}{260}       & 600      & \multicolumn{1}{c|}{$\sim$10\%} & \multicolumn{1}{c|}{$\sim$750}            & $\sim$150            & \multicolumn{1}{c|}{$\sim$24}          & $\sim$70\% \\ \cmidrule(l){2-9} 
\multicolumn{1}{|l|}{}                              & \multicolumn{1}{l|}{TON-IoT}      & \multicolumn{1}{c|}{260}       & 600      & \multicolumn{1}{c|}{$\sim$10\%} & \multicolumn{1}{c|}{$\sim$750}            & $\sim$150            & \multicolumn{1}{c|}{$\sim$24}          & $\sim$70\% \\ \cmidrule(l){2-9} 
\multicolumn{1}{|l|}{}                              & \multicolumn{1}{l|}{CIC IoT 2023} & \multicolumn{1}{c|}{260}       & 600      & \multicolumn{1}{c|}{$\sim$10\%} & \multicolumn{1}{c|}{$\sim$750}            & $\sim$150            & \multicolumn{1}{c|}{$\sim$24}          & $\sim$70\% \\ \midrule
\multicolumn{1}{|l|}{\multirow{4}{*}{TinyBERT}}     & \multicolumn{1}{l|}{X-IIoTID}     & \multicolumn{1}{c|}{55}        & 380      & \multicolumn{1}{c|}{$\sim$10\%} & \multicolumn{1}{c|}{$\sim$150}            & $\sim$30             & \multicolumn{1}{c|}{$\sim$5}           & $\sim$70\% \\ \cmidrule(l){2-9} 
\multicolumn{1}{|l|}{}                              & \multicolumn{1}{l|}{Edge-IIoTset} & \multicolumn{1}{c|}{55}        & 380      & \multicolumn{1}{c|}{$\sim$10\%} & \multicolumn{1}{c|}{$\sim$150}            & $\sim$30             & \multicolumn{1}{c|}{$\sim$5}           & $\sim$70\% \\ \cmidrule(l){2-9} 
\multicolumn{1}{|l|}{}                              & \multicolumn{1}{l|}{TON-IoT}      & \multicolumn{1}{c|}{55}        & 380      & \multicolumn{1}{c|}{$\sim$10\%} & \multicolumn{1}{c|}{$\sim$150}            & $\sim$30             & \multicolumn{1}{c|}{$\sim$5}           & $\sim$70\% \\ \cmidrule(l){2-9} 
\multicolumn{1}{|l|}{}                              & \multicolumn{1}{l|}{CIC IoT 2023} & \multicolumn{1}{c|}{55}        & 380      & \multicolumn{1}{c|}{$\sim$10\%} & \multicolumn{1}{c|}{$\sim$150}            & $\sim$30             & \multicolumn{1}{c|}{$\sim$5}           & $\sim$70\% \\ \midrule
\multicolumn{1}{|l|}{\multirow{4}{*}{Llama 3.2 1B}} & \multicolumn{1}{l|}{X-IIoTID}     & \multicolumn{1}{c|}{4600}      & 5500     & \multicolumn{1}{c|}{$\sim$10\%} & \multicolumn{1}{c|}{$\sim$15000}          & $\sim$3000           & \multicolumn{1}{c|}{$\sim$400}         & $\sim$70\% \\ \cmidrule(l){2-9} 
\multicolumn{1}{|l|}{}                              & \multicolumn{1}{l|}{Edge-IIoTset} & \multicolumn{1}{c|}{4600}      & 5500     & \multicolumn{1}{c|}{$\sim$10\%} & \multicolumn{1}{c|}{$\sim$15000}          & $\sim$3000           & \multicolumn{1}{c|}{$\sim$400}         & $\sim$70\% \\ \cmidrule(l){2-9} 
\multicolumn{1}{|l|}{}                              & \multicolumn{1}{l|}{TON-IoT}      & \multicolumn{1}{c|}{4600}      & 5500     & \multicolumn{1}{c|}{$\sim$10\%} & \multicolumn{1}{c|}{$\sim$15000}          & $\sim$3000           & \multicolumn{1}{c|}{$\sim$400}         & $\sim$70\% \\ \cmidrule(l){2-9} 
\multicolumn{1}{|l|}{}                              & \multicolumn{1}{l|}{CIC IoT 2023} & \multicolumn{1}{c|}{4600}      & 5500     & \multicolumn{1}{c|}{$\sim$10\%} & \multicolumn{1}{c|}{$\sim$15000}          & $\sim$3000           & \multicolumn{1}{c|}{$\sim$400}         & $\sim$70\% \\ \midrule
\multicolumn{1}{|l|}{\multirow{4}{*}{TinyT5}}    & \multicolumn{1}{l|}{X-IIoTID}     & \multicolumn{1}{c|}{60}        & 400      & \multicolumn{1}{c|}{$\sim$10\%} & \multicolumn{1}{c|}{$\sim$60}             & $\sim$12             & \multicolumn{1}{c|}{$\sim$2}           & $\sim$70\% \\ \cmidrule(l){2-9} 
\multicolumn{1}{|l|}{}                              & \multicolumn{1}{l|}{Edge-IIoTset} & \multicolumn{1}{c|}{60}        & 400      & \multicolumn{1}{c|}{$\sim$10\%} & \multicolumn{1}{c|}{$\sim$60}             & $\sim$12             & \multicolumn{1}{c|}{$\sim$2}           & $\sim$70\% \\ \cmidrule(l){2-9} 
\multicolumn{1}{|l|}{}                              & \multicolumn{1}{l|}{TON-IoT}      & \multicolumn{1}{c|}{60}        & 400      & \multicolumn{1}{c|}{$\sim$10\%} & \multicolumn{1}{c|}{$\sim$60}             & $\sim$12             & \multicolumn{1}{c|}{$\sim$2}           & $\sim$70\% \\ \cmidrule(l){2-9} 
\multicolumn{1}{|l|}{}                              & \multicolumn{1}{l|}{CIC IoT 2023} & \multicolumn{1}{c|}{60}        & 400      & \multicolumn{1}{c|}{$\sim$10\%} & \multicolumn{1}{c|}{$\sim$60}             & $\sim$12             & \multicolumn{1}{c|}{$\sim$2}           & $\sim$70\% \\ \bottomrule
\end{tabular}%
}
    \vspace{+0.1em}
    \caption{Performance results for Jetson nano and rasp pi 3  }
    \label{tab:device}
\end{table*}

%

% \usepackage{booktabs}
% \usepackage{multirow}
% \usepackage{graphicx}
\begin{table*}[]
\resizebox{\textwidth}{!}{%
\begin{tabular}{@{}|l|c|c|c|c|l|c|ccc|@{}}
\toprule
\multirow{3}{*}{\textbf{Model}} & \multirow{3}{*}{\textbf{\begin{tabular}[c]{@{}c@{}}Model Total \\ FLOPs \\ (GFLOPs)\end{tabular}}} & \multirow{3}{*}{\textbf{\begin{tabular}[c]{@{}c@{}}Model \\ weigths\\ (MB)\end{tabular}}} & \multirow{3}{*}{\textbf{\begin{tabular}[c]{@{}l@{}}Input\\ Tensors \\ (Billion)\end{tabular}}} & \multirow{3}{*}{\textbf{\begin{tabular}[c]{@{}l@{}}Intermediate\\ Activations\\ (MB)\end{tabular}}} & \multirow{3}{*}{\textbf{\begin{tabular}[c]{@{}l@{}}Output\\ Tensors\\ (MB)\end{tabular}}} & \multirow{3}{*}{\textbf{\begin{tabular}[c]{@{}c@{}}RAM \\ usage \\ (MB)\end{tabular}}} & \multicolumn{3}{c|}{\textbf{Latency (Classification Time)}} \\ \cmidrule(l){8-10} 
 &  &  &  &  &  &  & \multicolumn{1}{l|}{\multirow{2}{*}{Raspb Pi 3}} & \multicolumn{1}{l|}{\multirow{2}{*}{Jetson CPU}} & \multicolumn{1}{l|}{\multirow{2}{*}{Jetson GPU}} \\
 &  &  &  &  &  &  & \multicolumn{1}{l|}{} & \multicolumn{1}{l|}{} & \multicolumn{1}{l|}{} \\ \midrule
distilGPT2 & 7.25 & 327.65 & 4096.00  & 18.87 & 205.85 & 750 & \multicolumn{1}{c|}{24.16s} & \multicolumn{1}{c|}{724.78ms} & 144.96ms \\ \midrule
distilBERT & 7.25 & 265.45 & 4096.00  & 18.87 & 125.02 & 600 & \multicolumn{1}{c|}{24.16s} & \multicolumn{1}{c|}{724.78ms} & 144.96ms \\ \midrule
TinyBERT & 1.38 & 57.40 & 4096.00  & 5.11 & 125.02 & 380 & \multicolumn{1}{c|}{4.60s} & \multicolumn{1}{c|}{138.02ms} & 27.60ms \\ \midrule
Llama-3.2-1B & 137.44 & 4943.26 & 4096.00  & 134.22 & 525.34 & 5500 & \multicolumn{1}{c|}{458.13s} & \multicolumn{1}{c|}{13.74s} & 2.75s \\ \midrule
TinyT5 & 0.54 & 62.28 & 4096.00  & 4.19 & 131.60 & 400 & \multicolumn{1}{c|}{1.79s} & \multicolumn{1}{c|}{53.69ms} & 10.74ms \\ \bottomrule
\end{tabular}%
}
\vspace{+0.1em}
\caption{Theoretical performance of the SLM models deployed on the device}
\label{tab:teoricalPerformances}
\end{table*}

We exploited the formulas retrieved from the literature and listed in Table in \ref{tab:metrics} to calculate the theoretical performance for the model deployed on the device.
Considering the model characteristics listed in Table \ref{tab:LLMsmodels} and the PyTorch Overhead of 100 MB, we calculate performance for model size, RAM usage, and latency (classification time) as depicted in \ref{tab:teoricalPerformances}.
In the table, model weights refer to the parameters learned during training, stored as tensors, and determine the model's size.
The input tensor represents the data fed into the model.
Intermediate activations store temporary results during forward propagation through the model’s layers.
The output tensor contains the model's logits.
In contrast, Table \ref{tab:device}, shows the results of our experiments.

%discussion
Experimental results are aligned with the expected theoretical results.
The results show that all models require high CPU usage ($\sim$ 70\%) on both devices.
LLama results are the most demanding, as expected, due to the model's complexity.
The size and RAM usage remain the same for each device.
As expected,  tiny models require fewer resources for size and RAM usage than the distilled version, with TinyT5 as the smallest.
Similar behaviour is shown for the classification time, with TinyT5 being the fastest.
The models perform better on the Jetson Nano rather than the Raspberry for both classification time (-23 sec) and CPU usage (-60\%), thanks to GPU support and better hardware.
Overall, TinyBERT and TinyT5 show the best performance, making them the best choice for the edge scenario.
The Jetson Nano's superior performance makes it a better choice for edge nodes, where models that require higher resources can be deployed.
For example, in the case of Llama, which requires at least 5 GB of RAM, the Jetson Nano would be the ideal device thanks to its 8 GB RAM.

\section{Conclusions}\label{sec:conclusions}

In recent years, malware's growing sophistication and adaptability have highlighted the urgent need for innovative detection methods, particularly in resource-constrained edge computing environments. Traditional approaches have proven insufficient in addressing the challenges posed by modern malware, necessitating the exploration of advanced techniques. Lightweight Large Language Models (LLMs) have emerged as promising candidates for enhancing malware detection in edge scenarios. %These models, leveraging techniques like distillation and pruning, demonstrate an optimal balance between computational efficiency and detection accuracy, making them viable for deployment on edge devices.

In this paper, we tackled the issue of deploying LLMs in edge environments, investigating through a comprehensive experimental evaluation and careful design choices several challenges, such as maintaining high accuracy, mitigating overfitting, and addressing the computational constraints of edge devices.
To address inherent limitations of LLMs applied in an edge scenario,  we proposed an SML-based malware detection approach, where different edge devices in the architecture are fine-tuned to detect specific types of malware, allowing specialized and targeted threat identification.
By considering several available datasets, we evaluated the effectiveness of this architecture by considering the most popular lightweight LLMs,  and highlighted their strengths in detecting malware while adapting to diverse scenarios. 

Future research will focus on refining the design of our architecture, exploring advanced regularization techniques to mitigate biases, and optimizing deployment strategies for real-world edge environments. We also plan to complement our architecture with a central model to improve the detection performance of new threats by specialized edge devices along the lines presented in \cite{rondanini2024large}.
The central model will enable the integration of feedback from diverse edge devices to improve overall threat detection accuracy and maintain consistency in response strategies across the network.

\section*{Acknowledgment}
%The work by B. Carminati, E. Ferrari and C. Rondanini has been partially supported by CISCO under the research project:  ''Malware Detection for Edge-based Computing'' (Id 88749009).
This research work has been partially supported by Cisco Research under the research grant award:  ''Malware Detection for Edge-based Computing'' (Id 88749009).

\bibliographystyle{IEEEtran} 
\bibliography{biblio}

\end{document}